\documentclass[12pt,preprint]{aastex}

\usepackage{graphicx}   

\shorttitle{Infrared Observations of XRF~060218/SN~2006aj}
\shortauthors{Kocevski et al.}

\begin{document}

\title{Multicolor Infrared Observations of SN~2006aj, the Supernova Associated with XRF~060218 - Paper I}

\author{Daniel Kocevski \altaffilmark{1},  Maryam Modjaz \altaffilmark{2}, Joshua S. Bloom \altaffilmark{1}, Ryan Foley  \altaffilmark{1}, Daniel Starr \altaffilmark{3}, Cullen H. Blake \altaffilmark{2}, Emilio E. Falco  \altaffilmark{2}, Nathaniel R. Butler \altaffilmark{1}, Mike Skrutskie \altaffilmark{4},  Andrew Szentgyorgyi \altaffilmark{5} }

\altaffiltext{1}{Astronomy Department, University of California, 601 Campbell Hall, Berkeley, CA 94720 }
\altaffiltext{2}{Department of Astronomy, Harvard University, 60 Garden St., Cambridge, MA 02138}
\altaffiltext{3}{Gemini Observatory, Hilo, HI 96720}
\altaffiltext{4}{Department of Astronomy, P.O. Box 3818, University of Virginia, Charlottesville, VA 22903-0818.}
\altaffiltext{5}{Harvard-Smithsonian Center for Astrophysics,  Cambridge, MA 02138}

\email{kocevski@berkeley.edu, mmodjaz@cfa.harvard.edu, jbloom@astro.berkeley.edu, froley@astro.berkeley.edu, dstarr1@gmail.com, cblake@cfa.harvard.edu, efalco@cfa.harvard.edu, nat@astro.berkeley.edu, mfs4n@virginia.edu, saint@cfa0.cfa.harvard.edu}


\begin{abstract}
We report simultaneous multicolor near-infrared (NIR) observations of the supernova associated with x-ray Flash 060218 during the first 16 days after the high energy event.  We find that the light curve rises and peaks relatively fast compared to other SN Ic, with the characteristic broad NIR peak seen in all three bands.  We find that the rise profile before the peak is largely independent of NIR wavelength, each band appearing to transition into a plateau phase around day 10--13.  Since the light curve is in the plateau phase when our observations end at day 16, we can only place limits on the peak absolute magnitudes, but we estimate that SN~2006aj is one of the lowest NIR luminosity XRF/GRB associated SNe observed to date. The broad peaks observed in the {\em JHK$_s$} bands point to a large increase in the NIR contribution of the total flux output from days 10--16.  This evolution can be seen in the broad color and SED diagrams constructed using {\em UBVRIJHK$_s$} monochromatic flux measurements for the first 16 days of the event.  Ultimately, a 10-day rise time would make SN~2006aj an extremely fast rise SN Ic event, faster than SN~1998bw and SN~2003dh, which combined with its underluminous nature, indicates a lower amount of $^{56}$Ni ejected by the progenitor compared to other XRF/GRB-SNe.  Furthermore, the lack of significant color change during the rise portion of the burst points to little or no spectral evolution over the first 10 days of activity in the NIR.  We note that the onset of the plateau phase seen in the NIR roughly coincides with the reports of circularization of the SN ejecta indicating that the fast rise time and subsequent light curve behavior may be connected to the asphericity of the explosion.
\end{abstract}


 \maketitle


\section{Introduction} \label{sec:Introduction}
 
There is now a growing body of evidence to suggest that most long duration, soft-spectrum gamma-ray bursts (GRBs) are associated with the core collapse of massive stars.  This connection had long been suggested on theoretical grounds (Colgate 1968, Woosley 1993) as well as due to the similarities between GRB and SNe energetics.  As of early 2006, this connection had been solidified by the detection of a total of three GRBs (GRB 980425/SN 1998bw, GRB 031203/SN 2003lw , and GRB 030329/SN 2003dh) with direct spectroscopic connections to supernova events.  The early optical spectra of these supernovae components for which high-quality spectra exist show remarkable similarities, typically deficient of hydrogen and helium lines, a property consistent with Type Ic supernova.  Furthermore, very broad absorption lines of \ion{O}{1}, \ion{Ca}{2}, and \ion{Fe}{2} indicate high expansion velocities and a kinetic energy ($E_{K}$) going much higher than the values inferred in typical SN Ic events.  Despite great interest in understanding the GRB-SN connections, only a few events have been observed well in the infrared \citep{Lipkin04,Cobb06,Bloom04}.  Since estimates of $E_{K}$ and the amount of $^{56}$Ni produced in the explosion depend on the total bolometric output, IR observations become particularly important for our understanding of the fundamental properties of these events.

On 2006 February 18 at 03:34:30 {\sc UTC} the Swift spacecraft detected and localized XRF~060218 \citep{Cusumano06a}.  The burst was immediately recognized as an unusual event when compared to other bursts observed by Swift.  It was far longer than any GRB or XRF previously detected by the spacecraft, with a $T_{90}$ duration of roughly 2000 seconds.  It also exhibited a soft gamma-ray spectrum \citep{Barbier06} and an unusual afterglow that brightened for the first ten hours after the explosion before transitioning to a more common power law decay \citep{Cusumano06b,Campana06}.  The first indications of an underlying SN were noted by spectra taken roughly 3 days after the event \citep{Masetti06} followed by reports of a rebrightening of the optical counterpart shortly thereafter \citep{DAvanzo06}.  Like previously observed XRF/GRB-SN the spectra of SN~2006aj showed continuum and broad-line emission with significant amounts of \ion{Fe}{2} and \ion{Si}{2} $\lambda 6355$ but no H or He emission.  The SN~2006aj spectra best matches those of SN~2002ap and SN~1997ef at similar epochs, both broad line SN Ic events but neither of which have been conclusively shown to be associated with a XRF or GRB \citep{Modjaz06}.  Emission line measurements by \citet{Mirabal06a} gave a redshift of $z = 0.0331$ making it one of the closest XRF/GRB ever observed, second only to GRB~980425.  Such a close distance, $\sim 140$ Mpc, meant that XRF~060218 was similar to two other nearby XRF/GRB-SN events, GRB~980425 and GRB~031203, in that they were all significantly underluminous in the total amount of emitted gamma-ray, x-ray, and radio energies when compared to the majority of cosmological bursts \citep{Soderberg06}.  The {\em V} band peak of SN~2006aj occurred at roughly T = 10.0 days (9.7 days in the rest frame) with a peak apparent magnitude of $M_{v} = 18.7 \pm 0.2$ mag \citep{Modjaz06,Sollerman06,Mirabal06a}, making it both the fastest evolving and least luminous of the XRF/GRB-SN observed to date.  Pre-burst imaging of the SN~2006aj field by SDSS provided its host galaxy magnitudes \citep{Cool06,Hicken06}, which given its distance reveal a low-luminosity dwarf galaxy as the host, with low metallicity \citep{Modjaz06,Wiersema07}, similiar to the majority of other GRB host galaxies \citep{Fruchter06, Stanek06,Kewley06}.

PAIRITEL observations of XRF~060218/SN~2006aj started on Feb 20th, well before the observed {\em V} band and {\em J} band peaks reported by \citet{Modjaz06} and \citet{Cobb06}, and continued for the next 14 days.  Here in Paper I we report on the light curve properties and host-subtracted absolute magnitudes of SN~2006aj in the  {\em J}, {\em H}, and {\em K$_s$} infrared bands during these epochs.  We reserve a more detailed presentation of the broadband SED evolution during this period for Paper II \citep{Modjaz07}.  We present our observation and data reduction techniques and results in $\S 2$ and discuss the implications of our observations to the further understanding of the XRF/GRB-SN phenomena in $\S 3$.

\section{Data $\&$ Analysis} \label{sec:Data}

All of the infrared data presented in this paper were taken with the fully
automated PAIRITEL project \footnote{http://www.pairitel.org} located on Mt. Hopkins in Arizona.  The
1.3m telescope contains the three Near-Infrared Camera and Multi-Object
Spectrometer (NICMOS3) arrays formerly of the 2MASS project
\citep{Skrutskie06} to simultaneously read near infrared {\em J}, {\em H}, and {\em K$_s$}
(1.2, 1.6, and 2.2 $\mu$m respectively) band images.  Each image
consists of a 256 x 256 array with a pixel scale of 2 arcsec/pixel and
an integration time of 7.8 s per image.  These individual images are
dithered in order to correct for bad pixels, and mosaics are created
by drizzling the images, producing a Nyquist-sampled image with a
pixel scale of ~1 arcsec/pixel. The telescopeÕs entire operating
software has been built on the Python programming language and allows
for autonomous operation which can be monitored and controlled
remotely \citep{Bloom05} making it excellent for the followup of
transient events.

Observations of XRF 060218 began on the morning of Feb 20th and
continued until the object fell below our airmass limits some 16 days
later.  A total of $>9000$ individual images were taken in the
{\em J}, {\em H}, and {\em K$_s$} wavebands to produce a total of 96 reduced mosaics.  The
individual images were dithered through a predetermined pattern for a
single epoch of observations and then reduced through a custom
pipeline. Bias and sky frames for each image were produced by median-combining several dithered exposures before and after the frame, whereas archival frames were used for the bad-pixel masks and flat
fields.  The final mosaics were produced using a drizzle technique
\citep{Fruchter97} for each epoch and filter with an equivalent
exposure time ranging between 2 and 30 minutes.  The average exposure
time of each mosaic generally increased as the object moved to higher
air mass in subsequent nights.  Of the 96 mosaics produced through the
automated reduction pipeline, 3 were unusable due to poor
transmission on three nights, leaving $\sim$ 31 usable mosaics per filter.  

A custom pipeline was used to perform photometry of the final {\em J}, {\em H}, and {\em K$_s$}
mosaics, using aperture photometry via the {\em Sextractor} package \citep{Bertin96} and PSF photometry via the {\em NSTAR} routine in IDL to estimate the instrumental magnitudes of every object in each mosaic.  These values were then compared to the original 2MASS catalog of the same field for zeropoint determination of each reduced frame.  Finally, an IDL routine was employed to produce a light curve for every stellar object in the field for a final relative calibration.  The resulting median $\Delta m$ variations in these stellar light curves allow for additional corrections to be made to the SN light curve that account for any errors in the zeropointing of the individual mosaics.

Overall, we find that the aperture and PSF fitting routines generally yield equivalent values for {\em J}, {\em H}, and {\em K$_s$} band mosaics. We ultimately chose to use aperture photometry with a floating aperture radius roughly equal to 1.5, 1.75, and 2.0 times the seeing (FWHM) for the {\em J}, {\em H}, and {\em K$_s$} bands respectively.  These factors were empirically determined to maximize the signal to noise in each band and result in aperture radii roughly ranging from 3 to 5 pixels.

To account for the host galaxy contribution in the SN light curve, we employ a modified version of the ISIS image subtraction routine developed by \citet{Alard00}.  We revisited the XRF~060218 field nine months after our original observations to obtain deep template imaging of the host galaxy when the SN contribution was negligible.  In all, we obtained an effective exposure of roughly 5.5 hours in the {\em J}, {\em H}, and {\em K$_s$} bands after co-adding several nights of imaging of the host galaxy.  Using this late-time template, we were able to produce residual images containing the measured flux contribution minus the host galaxy, leaving only the SN flux.  In order to quantify the error involved in this subtraction process, we inserted a series of fake stars into the original mosaics and tracked the changes in their known magnitude through each subtraction.  The resulting variations in the fake star light curves are then accounted for in the final host subtracted SN light curve.  We find that the host galaxy's flux contribution to the uncorrected SN light curve in most of the subtractions is roughly equal to the host magnitude found in the late time template imaging, indicating that the host galaxy can be treated as a point source at the resolution of our images.  This allows us to subtract the host contribution in catalog space in order to reduce the scatter in the final SN light curve.

\section{Results} \label{sec:Results}

Figure 2 shows our host-subtracted {\em JHK$_s$} light curves for SN~2006aj over the course of the first 16 days since the burst plotted in the observer frame, with the {\em H} and {\em K$_s$} bands shifted by 0.9 and 1.5 mag respectively for clarity. Each data point reflects the median average of all observations taken in a single night of observations.  The error bars represent the 1 $\sigma$ standard deviation about this median magnitude added in quadrature to the individual photometric measurement errors of a single mosaic.  On nights in which only one mosaic was available, only the photometric error was used, which is most likely an underestimate of the true (statistical + systematic) error.

The characteristic broad peak of SN Ic events can be seen in all three bands, where the light curves flatten between day 10 and 13, with the {\em K$_s$} band peak being more uncertain due to a gap in the available data. \citet{Cobb06} has reported on {\em J}  band data peaking around day 17, just beyond our observability range, so we may miss the peak magnitude in {\em J} by a day.  To estimate the peak magnitudes of our data set, we employ a spline fit to the light curve of each individual filter and measure apparent peak magnitudes of {\em J}: 16.76 $\pm$ 0.03,  {\em H}: 16.65 $\pm$ 0.05, and {\em K$_s$}: 16.39 $\pm$ 0.21.  Our estimate of the {\em J} band peak is consistent with the peak {\em J} band measurements made by \citet{Cobb06} of the the SN and host of 16.65 $\pm$ 0.06 mag at day 17, one day after our last observation.  Both the {\em H} and {\em K$_s$} bands are expected to peak after the {\em J} band, so the above values can only act as upper limits to the {\em HK$_s$} peak magnitudes.  All three PAIRITEL bands appear to transition to the plateau phase at roughly the same time, with the {\em J} band possibly preceding the {\em H} and {\em K$_s$} bands.  The Galactic line-of-sight extinction values for each of the bands, $A_{J} = 0.112$, $A_{H} = 0.071$, $A_{K_s} = 0.045$ \citep{Schlegel98}, have been subtracted from the apparent peak magnitudes.  Extinction from the host galaxy has been estimated by \citet{Guenther06} through the measurements of the equivalent width (EW) of the Na I D lines along the line of sight.  They find that most of the extinction is attributable to our Galaxy with a visual Galactic extinction of $A_{V}$ = 0.39 $\pm$ 0.02 mag and a corresponding visual host frame extinction of $A_{V}$ = 0.13 $\pm$ 0.01 mag.  If we assume a ratio of total-to-selective extinction of $R_{V}$ = 3.1 \citep{ccm89}, we estimate near infrared extinction values from the host galaxy of roughly $A_{J}$ = 0.035.,  $A_{H} = 0.022$, $A_{K_s} = 0.015$ mag, which is considerably less than the extinction due to the line of sight through our own Galaxy.  Because the well-studied hosts \citep{vrees04,jakob03,sNf04,watson06,Butler06} of both nearby and cosmological GRBs appear to have a lower dust-to-metal ratio than the Milky Way and probably flatter extinction laws, we also consider total-to-selective extinction values of $R_{V} = 2.0$ (5.0), which yield roughly $A_{J} = 0.035 (0.035)$,  $A_{H} = 0.024 (0.022)$, $A_{K_s} = 0.015 (0.015)$ mag.  None of these values deviates significantly from the $R_{V} = 3.1$ assumption and the effects on the final light curves are within the error of our photometry.  At a distance of $\sim 141$ Mpc ($z = 0.0335$, $H_{0} = 72$ km s$^{-1}$, $\Omega_{m} = 0.3$, $\Omega_{\Lambda} = 0.7$) the peak absolute magnitude in {\em J} and the upper limit in {\em H} and {\em K$_s$}, taking into account both sets of extinction values, comes to {\em J}: $-$19.02 $\pm$ 0.03, {\em H}: $-$19.13 $\pm$ 0.05, and {\em K$_s$}: $-$19.39 $\pm$ 0.24.  In this case, these values are not k-corrected due to the low redshift of the event.  Photometry from the deep late time images gives the host galaxy's {\em JHK$_s$} apparent magnitudes {\em J}: 18.99 $\pm$ 0.16, {\em H}: 18.52 $\pm$ 0.22, and {\em K$_s$}: 18.73 $\pm$ 0.34, uncorrected for Galactic extinction.  This represents 33.6$\%$, 36.7$\%$, and 27.8$\%$ of the {\em J}, {\em H}, and {\em K$_s$} flux contribution to the SN light curve at early times, respectively.

It should be noted that fits to the early X-ray data \citep{Butler07} collected by the Swift spacecraft show evidence for excess X-ray absorption beyond that due to our Galaxy alone $N_{H,{\rm Galactic}}=1.11 \times 10^{21}$ cm$^{-2}$ \citep{dickey1990}: $N_H = 2.9 \pm 0.5 \times 10^{21}$ cm$^{-2}$,  at $z=0.033$. Galactic extinction \citep{ccm89} and the Galactic $A_V$--$N_H$ relation \citep{pns95} would then imply the following absorption magnitudes in addition to the absorption by the Galaxy: $A_{J}$= 0.51 $\pm$ 0.07, $A_{H}$= 0.33 $\pm$ 0.04, $A_{K_s}$= 0.21 $\pm$ 0.03. These extinction values disagree with the values calculated through the use of the EW of the Na I D by \citet{Guenther06} and highlights the uncertainty in applying the host galaxy extinction correction.  Thus the X-ray inferred results are not subtracted from our photometry stated above but should be considered as upper limits to the possible extinction from the host galaxy.  We briefly discuss possible explanations to these extinction disagreements in $\S 3$.


We compare the observer frame NIR light curve to the light curve behavior at optical wavelengths in Figure 3.  Here we plot our {\em JHK$_s$} photometry along with {\em UBVRI} measurements made by \citet{Mirabal06b} for the first 26 days of the event using the 1.3m and 2.4m MDM telescopes.  Each of the individual light curves is connected with a cubic spline for clarity. The final \citet{Mirabal06b} photometry that we adopt for our light curve comparison is host-subtracted and extinction-corrected assuming apparent host magnitudes of $U = 20.10$, $B = 20.41$, $V = 20.09$, $R = 19.91$, $I = 19.54$ and galactic extinction of $A_{U} = 0.77$, $A_{B} = 0.61$, $A_{V} = 0.47$, $A_{R} = 0.38$, and $A_{I} = 0.28$.  The previously observed trend of broader light curves at longer wavelengths can be seen when comparing the turn over time in the shorter wavelengths compared to our {\em JHK$_s$} measurements. The detailed modeling of the light curve behavior and the time-dependent radiative transfer calculations required for a full understanding of this trend have been accomplished for the case of SNe Ia \citep{Kasen06}, but have yet to be done for SNe Ib/c and are beyond the scope of this paper.  This increase in the NIR flux contribution at late times can be nicely seen in the color diagram shown in Figure 4.  Here we display our host subtracted and extinction corrected {\em J} band photometry minus the {\em UBVRI} data from \citet{Mirabal06b} and our {\em HK$_s$} PAIRITEL measurements.  The gradual steepening of the color differences is a function of $\Delta \lambda_{ \rm eff}$ between the two filters and quantifies the relative difference between the light curve decay rates at shorter wavelengths with respect to the NIR.  When fit with a linear function such that $\Delta{m} = \Delta{m_0}-s t$, the individual color difference slopes $s$ in Figure 4 are: {\em J$-$K$_s$}: $-$0.020 $\pm$ 0.012, {\em J$-$H}: $-$0.011 $\pm $0.008, {\em J$-$I}: $-$0.014 $\pm$ 0.006, {\em J$-$R}: $-$0.032 $\pm$ 0.005, {\em J$-$V}: $-$0.048 $\pm$ 0.005, {\em J$-$B}: $-$0.097 $\pm$ 0.0046, {\em J$-$U}: $-$0.143 $\pm$ 0.007 mags/day.

The color evolution properties can also be seen in the spectral energy distribution (SED) shown in Figure 5 constructed using this same broad band {\em UBVRIJHK$_s$} photometry .  We converted the {\em UBVRIJHK$_s$} photometry to monochromatic flux values using Johnson-Morgan $f_{\nu,\rm eff}$ flux zeropoints from \citet{Fukugita95} for the \citet{Mirabal06b} data set and the 2MASS $f_{\nu,\rm eff}$ flux zeropoints given in \citet{Cohen03} for our PAIRITEL data. The figure shows several SED curves ranging from roughly 1.5 to 12 days after the Swift trigger as measured in the observer frame of the host galaxy; the individual SEDs are connected with a cubic spline for clarity.  From the plot it can be seen that the shorter wavelength contribution to the bolometric flux of the SN emission is much higher near the onset of the event and steadily decreases after day 7 in the host frame (day $\sim$ 9 in the observer frame).  As a result, a broadening of the SED profile can also be seen at late times as the NIR contribution to the bolometric flux increases, consistent with the conclusions drawn from Figure 4.  A much more detailed analysis and discussion of the evolution of the XRF~060218/SN~2006aj SED using our PAIRITEL observations in conjunction with optical {\em UBVr'i'} photometry taken with the Mt Hopkins 48-inch will be covered in Paper II (Modjaz et al., in prep).

It is difficult to quantify the amount, if any, of the infrared emission over the first 16 days that is a result of light echos, a scenario in which the dust in the vicinity of the burst progenitor absorbs and then reradiates the optical and UV emission from the explosion.  This would occur at a distance greater than the radius $R_{c}$ at which the dust grains are destroyed due to sublimation, which when assuming a peak luminosity which is 70$\%$ of 1998bw gives a rough $R_{c}$ $\sim$ 8-9 pc \citet{Waxman00}.  If the dust at this distance has a characteristic equilibrium temperature of $\sim$2300 K (the temperature above which the grains are destroyed), then the resulting reradiated light would peak at $\nu \sim 2(1+z) \mu$m, where $z$ is the redshift of the event, which at low $z$ corresponds roughly to the {\em K} band \citet{Reichart01}.  Although the prompt UV and optical emission along the line of sight could contribute infrared emission on any time scale, the majority of the reradiated light is expected to be delayed as emission arrives from higher latitudes.  We conclude that this effect, if present, contributes  very little to our overall photometry primarily because the large sublimation radius would make the peak in the reradiated light occur beyond our last observation on day 16.  Furthermore, a time varying excess in the {\em K$_s$} would be expected if this effect were significant, which is not reflected in the SED.

\section{Discussion} \label{sec:Discussion}

There is no significant evidence that any of the individual NIR {\em JHK$_s$} light curves observed with PAIRITEL peaks significantly faster than the others, although the {\em H} and {\em K$_s$} light curves are expected to peak after the {\em J}  band which would occur outside of our observation window. Furthermore, the rise profile in the individual {\em J}, {\em H} and {\em K$_s$} bands before the peak appears to be largely independent of wavelength, with the light curve in each filter exhibiting roughly the same slope. This would make the chromatic rise time properties of SN~2006aj consistent with bolometric rise time results reported by \citet{Yoshii03} for SN~2002ap and points to little or no spectral evolution in the NIR in the days preceding peak brightness.  This can be seen in the color difference plot shown in Figure 5, which shows no significant trend in color evolution between the infrared wavelengths during the rise portion of the burst. Furthermore, the sharp {\em V} band peak reported by \citep{Modjaz06,Sollerman06,Mirabal06a} points to a relatively large increase in the NIR contribution to the total flux output after day $\sim$ 10. 

The disagreements between the X-ray and optically inferred extinctions are striking but not uncommon in the literature \citep[e.g.,][]{Campana06, Guenther06, Sollerman05, Wiersema07, Watson07}.  As mentioned above, GRB host galaxies tend to have lower dust to metals ratio than, and probably extinction laws that deviate from that of the Galaxy, possibly explaining the discrepancy between the Na I D and $N_H$ determined extinction values.  Furthermore, \citet{Watson07} has reported on a disagreement between the $N_H$ column densities as inferred from Ly$\alpha$ absorption to the metal column densities from soft X-ray absorption, which they speculate is because the two diagnostics do not probe the same gas environments.  Their results would imply that the $N_H$ column measured from the early X-ray data would have the effect of overestimating the extinction that would be inferred through the use of the $A_V$--$N_H$ relation, possibly explaining the disagreement described in S 3.  Therefore the exact host extinction could be the effect of a complex picture which is still not completely understood.  Fortunately, this disagreement on the host galaxy extinction affects our NIR data far less than those at shorter wavelengths highlighting the importance of NIR observations.

If we consider day 13 as being the onset of the NIR plateau phase of SN~2006aj, then the broad {\em J} band peak would occur several days after the peak in the {\em V} band, which is estimated at 9.7 days in the rest frame by \citet{Modjaz06}.  This would make our  {\em J} band rise time consistent with the previously observed trend in most SNe, in which the optical peak precedes the infrared peak, as was the case with GRB~980425/SN~1998bw and SN~2003lw, both of which peaked 1.6 \citep{Galama98} and 5 \citep{Malesani04} days later in {\em I} band than in {\em V} band respectively. The observed flattening of the NIR as early as day 10 makes SN~2006aj the fastest of the XRF/GRB-SNe observed to date, being substantially shorter in rise time than both GRB~980425/SN~1998bw and GRB~030329/SN~2003dh \citep{Galama98,Matheson03}.  The fast evolution and plateau seen in SN~2006aj is reminiscent of the light curve properties of broad-lined SNe Ic (SNe Ic BL), highly energetic core collapse SNe, that do not show likely associations with GRBs.  Two such events, SN~2002ap and SN~1997ef, have broad bolometric light curves where the latter is estimated to peak in $\sim$ 10$-$12 days, assuming the explosion date inferred by \citet{Mazzali02}.  Similiarly, our peak absolute magnitude of $-$19.02 $\pm$ 0.03 measured in {\em J} and $-$18.7 Measured in {\em V} \citep{Modjaz06,Sollerman06,Mirabal06a} would make SN~2006aj the faintest of the XRF/GRB associated SNe, yet still brighter than the average broad-lined SN Ic.  This can be seen in Figure 6, where the {\em J} and {\em R} band light curves of XRF~060218/SN~2006aj are plotted along the {\em J} and {\em R} band light curves of GRB~980425/SN~1998bw and SN~2002ap.  

Both the peak luminosity and rise time properties of SNe light curves are generally governed by the amount of $^{56}$Ni produced in the explosion, leading many authors to conclude that the fast evolving and underluminous nature of SN~2006aj points to smaller amounts of $^{56}$Ni compared to other XRF/GRB-SNe.  Modeling by \citet{Mazzali06a} estimates the $^{56}$Ni mass produced by SN~2006aj to be roughly $0.2M_{\odot}$, compared to SN~1998bw and SN~2003dh which are thought to have produced roughly 0.38$-$0.45M and 0.45$-$0.65M respectively \citep{Mazzali06b}.  Yet, this low estimate of $^{56}$Ni for SN~2006aj is still higher than the $\sim 0.09M_{\odot}$ estimated for the SN~2002ap SN Ic BL \citep{Foley03} which reflects the amount typical produced by normal
core-collapse SNe such as SN~1987A and SN~1994I.

SN~2006aj was also peculiar in several other aspects.  First, the measured peak spectral energy
of the gamma-ray emission $E_{pk} = 4.9$ keV, where $E_{pk}$ is the max of the $\nu F_{\nu}$ spectra and hence where most of the gamma-ray energy is radiated, is extremely soft.  This is in comparison to the GRB counterparts of SN~1998bw, 2003dh and 2003lw, which all had $E_{pk}$ values of roughly 55, 79, and 159 keV respectively. Furthermore, the modeling done by \citet{Mazzali06a} estimates an explosion kinetic energy of $E_{K} \approx 2 \times 10^{51}$ erg and total ejected mass of $M_{ej} \approx 2M_{\odot}$, both of which are lower than the values estimated for the other SNe with associated GRBs, but yet on the high end of the distribution of these values when compared to normal SN Ic events.  Considering these relatively low $E_{pk}$, $E_{K}$, and $M_{ej} $ values along with the low luminosity and fast rise time, it becomes clear that GRB~060218/SN~2006aj was indeed a peculiar type of event.  It is quite likely that SN~2006aj, with its low $E_{K}$ and $M_{ej}$ values, is an intermediate type of event.  Extreme in many ways when compared to broad-line core-collapse SNe like SN~2002ap and 1997ef which emit little or no high energy emission, but rather weak when compared to previously observed XRF/GRB-SNe. \citet{Li-Xin06} has recently quantified a previously observed trend that correlates a GRB's $E_{pk}$ value and the peak luminosity of its SNe emission, and hence the produced $^{56}$Ni mass.  Using this correlation, he estimates that the $E_{pk}$ values for SN~2002ap and 1997ef, if they had possessed an associated GRB, would have occurred in the UV, far below the range of what is considered a GRB or XRF.  This of course is based on the assuming this correlation is real and not an artifact of observational biases or source evolution.  All of the SN~2006aj properties discussed above place it between the previously observed GRB-SNe sample and these low-luminosity and low estimated $E_{pk}$ broad-lined events, possibly pointing to a continuous distribution of SN events that emit at increasingly higher energies. 

One possible explanation for such a continuous distribution of gamma ray energies is that the soft-spectrum and low-luminosity events like XRF 060218 are due to off-axis observations of typical long-duration GRB-SNe \citep{Yamazaki03}. Although this off-axis scenario was originally involked to explain the peculiar high energy properties of GRB-SNe like SN~1998bw, recent numerical simulations performed by \citet{Maeda06} have shown that the viewing angle of asymmetric explosions would also have an effect on the observed luminosity and time of peak of the associated SNe.  Through the use of 3D Monte Carlo simulations, the authors find that asymmetric SNe would appear to peak earlier when viewed at small $\theta$ from the z-axis of the explosion, this being due primarily to low and extended $^56$Ni densities in that direction.  The resulting luminosity as seen in the z-direction is then boosted correspondingly because the photons can only diffuse out in this direction.  This small $\theta$ scenario would then predict fast peaking, high luminosity, and high $E_{pk}$ events which is inconsistent with the parameter distribution seen in XRF 060218 which has a distinctly fast time to peak but a remarkably low luminosity and $E_{pk}$.  Another difficulty of the off-axis interpretation is that it would require a very large intrinsic total energy of roughly $\sim 10^{53}$ ergs, which would be inconsistent with the relatively low kinetic energy measurements made from radio observations \citep{Soderberg06} at late times when the jet collimation and/or SNe asymmetry should no longer be significant.  Furthermore, the probability of seeing a jet off-axis goes down significantly with increasing $\theta_{j}$ resulting in inconsistent rate ratios between high and low energy events as noted by \citet{Cobb06}. Therefore, although off-axis models can be made to successfully explain the observed range of GRB/XRF energetics, the range of SN properties such as $E_{K}$, $M_{ej}$, the measurements of which are generally independent of collimation, suggest that viewing angle effects alone cannot account for the full diversity of XRF/GRB-SNe properties.  

On the other hand, the distribution of the intrinsic properties of the progenitor star that would lead to such a continuous distribution in the observed XRF/GRB-SNe properties is also not entirely clear.  A qualitative correlation between the estimated progenitor mass and the resulting explosive energy seen among several broad-line and normal core-collapse SNe \citep{Mazzali02} points to a high initial stellar mass along with the progenitor's rotation rate, both of which would govern whether the jet associated with the GRB breaks out of the progenitor, as likely candidates.  Nonetheless, future broad-band observations of intermediate events like SN~2006aj and SN~2002ap may be crucial to the examination of this distribution. 

Finally, we note that the \citet{Gorosabel06} have reported evidence of a possible circularization of the explosion geometry, which like previous XRF/GRB-SN is thought to be aspherical at early times \citep{Mazzali01,Hjorth06}.  The authors show a rotation in the polarization angle of $\sim$90 degrees at 14 $\lesssim t \lesssim$ 39 days which they interpreted as a circularization of the previous highly aspherical geometry.  Similarly, \citet{Pian06} reported a change in the decay slope of the photospheric expansion velocities vs. time as determined through modeling of the spectra at the various epochs, which could also be interpreted as evidence for circularization.  This circularization time of $t \sim 14$ roughly corresponds with the onset of the plateau phase in our NIR observations of SN~2006aj.  One could speculate that the fast rise observed in SN~2006aj is a result of fast and aspherical outflow which becomes spherical after $t \sim 14$ days, creating the observed plateau phase.  Although, the fact that a general broadening of SN light curves in the NIR are seen in other types of (presumably spherical) SN events would point to a mechanism common to all SNe, such as radiative transfer and/or the ionization changes in the SN ejecta \citep{Kasen06}, rather than the geometry of the explosion.  Future NIR observations on the existence, or lack thereof, of a NIR plateau in normal (non-broadlined) SN Ic along with late-time spectral modeling and polarization measurements should shed light on any potential correlation between the expansion geometry an the resulting NIR light curve.

\section{Acknowledgments} \label{sec:acknowledgments}

The Peters Automated Infrared Imaging Telescope (PAIRITEL) is operated by the Smithsonian Astrophysical Observatory (SAO) and was made possible by a grant from the Harvard University Milton Fund, the camera loan from the University of Virginia, and the continued support of the SAO and UC Berkeley. Partial support for PAIRITEL operations and this work come from NASA grant NNG06GH50G ("PAIRTEL: Infrared Follow-up for Swift Transients"). This work was conducted under the auspices of a DOE SciDAC grant (DE-FC02-06ER41453), which provides support to J. S. B.'s group. J. S. B. thanks the Sloan Research Fellowship for partial support. D.K. acknowledges financial supported through the NSF Astronomy $\&$ Astrophysics Postdoctoral Fellowships under award AST-0502502.





\bigskip

\section*{Figure Captions}

{\bf Fig. 1.} -  Color composite finding charts of the host galaxy (seen in SDSS pre-imaging; left) and SN~2006aj (seen in PAIRITEL; right). The SDSS image is made using $r$, $i$, $z$-band images (Cool et al. 2006) and the IR image using $J$, $H$, $K_s$ from stacked mosaics obtained on 3 March 2006 UT, near in time to the peak of the optical emission of the SN. North is up and East is to the left.

{\bf Fig. 2} - The {\em J} band (blue circles), {\em H} band (green circles), and {\em Ks} band (red circles) aperture photometry light curve of XRF~060218/SN~2006aj after subtraction of the host contribution.  The {\em H} and {\em K$_s$} band light curves have been shifted by $-$0.9 and $-$1.5 magnitudes respectively for clarity.  The three light curves have been corrected for Galactic and host line of sight extinction. The error bars represent the quadrature sum of the photometric errors associated with all the mosaics produced in a single night of observations plus the 1 $\sigma$ standard deviation about their median magnitude.  A flattening of the light curves in all wavebands can be seen by Day 16.


{\bf Fig. 3} - Our near infrared PAIRITEL light curves plotted along with {\em UBVRI} data taken from \citet{Mirabal06b}. A steady progression of wider light curves at longer wavelengths is clearly seen.

{\bf Fig. 4} - Color diagram showing the subtraction of the PAIRITEL {\em J} band and {\em H} and {\em K$_s$} band data as well as the \citet{Mirabal06b} {\em UBVRI} photometry.  The slope of the color differences becomes steeper with increasing $\Delta\lambda_{ \rm eff}$, showing a large increase in the NIR contribution to the total bolometric flux output of the SN at later times.

{\bf Fig. 5} - A broad spectral energy distribution diagram produced by converting the {\em JHK$_s$} band data and the \citet{Mirabal06b} {\em UBVRI} photometry into monochromatic flux values and plotted vs the effective frequency of that band pass, corrected for the redshift of the host galaxy.  The NIR contribution to the total energy represented by the SED can be seen to increase at later times.

{\bf Fig. 6} - A comparison of the {\em R} and {\em J} band light curve properties of XRF~060218/SN~2006aj (PAIRITEL), GRB~980425/SN~1998bw \citep{Galama98}, and SN~2002ap \citep{Foley03,Yoshii03}.  We use the explosion date of January 28.9 UT (JD = 2452303.4) inferred by \citet{Mazzali02} as the $t=0$ time for SN~2002ap.

\bigskip


\section*{Figures}

\begin{figure}  \label{fig:finder}
\plotone{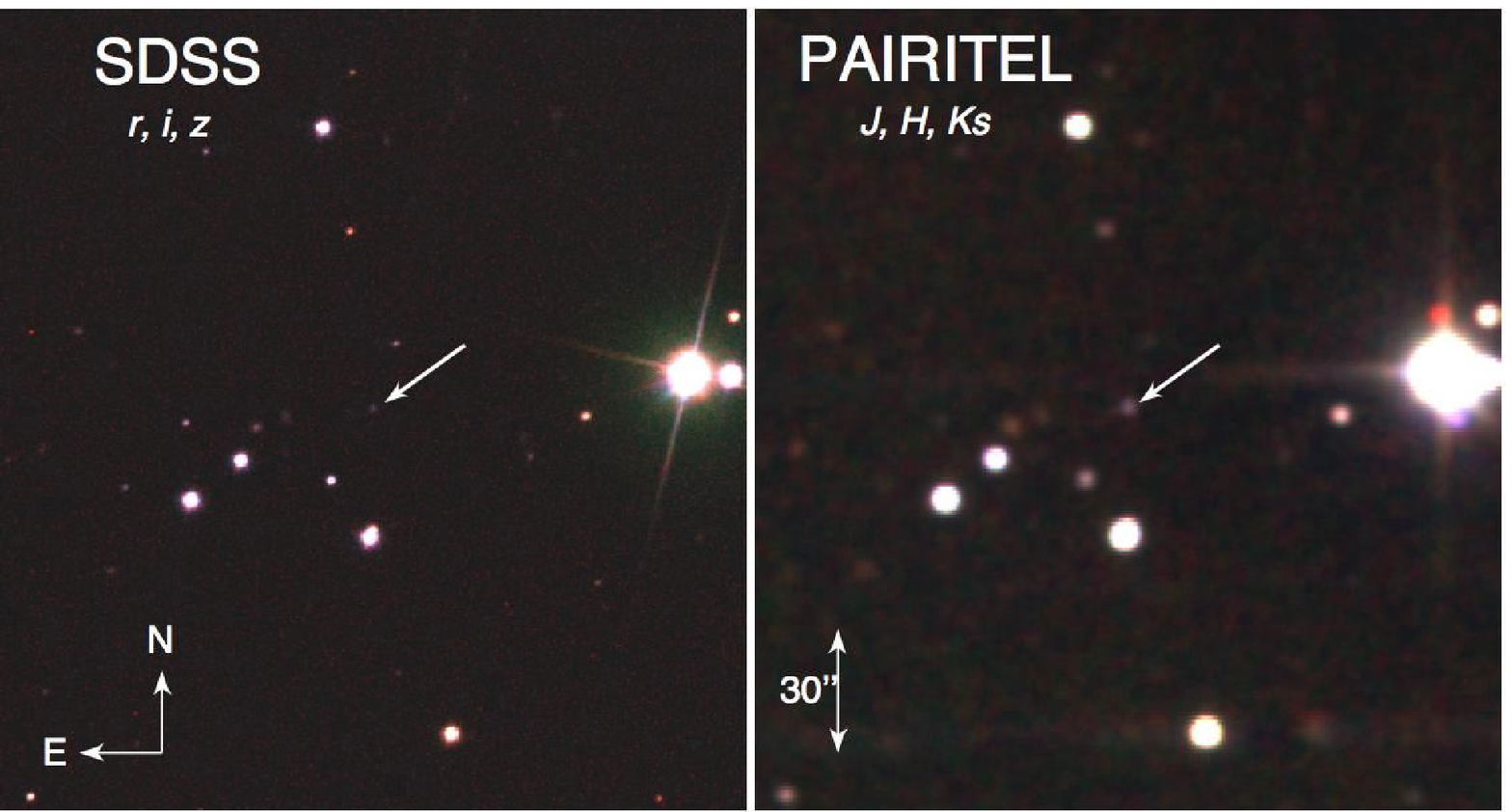}
\end{figure}

\begin{figure}  \label{fig:jhk-lightcurve-shift}
\plotone{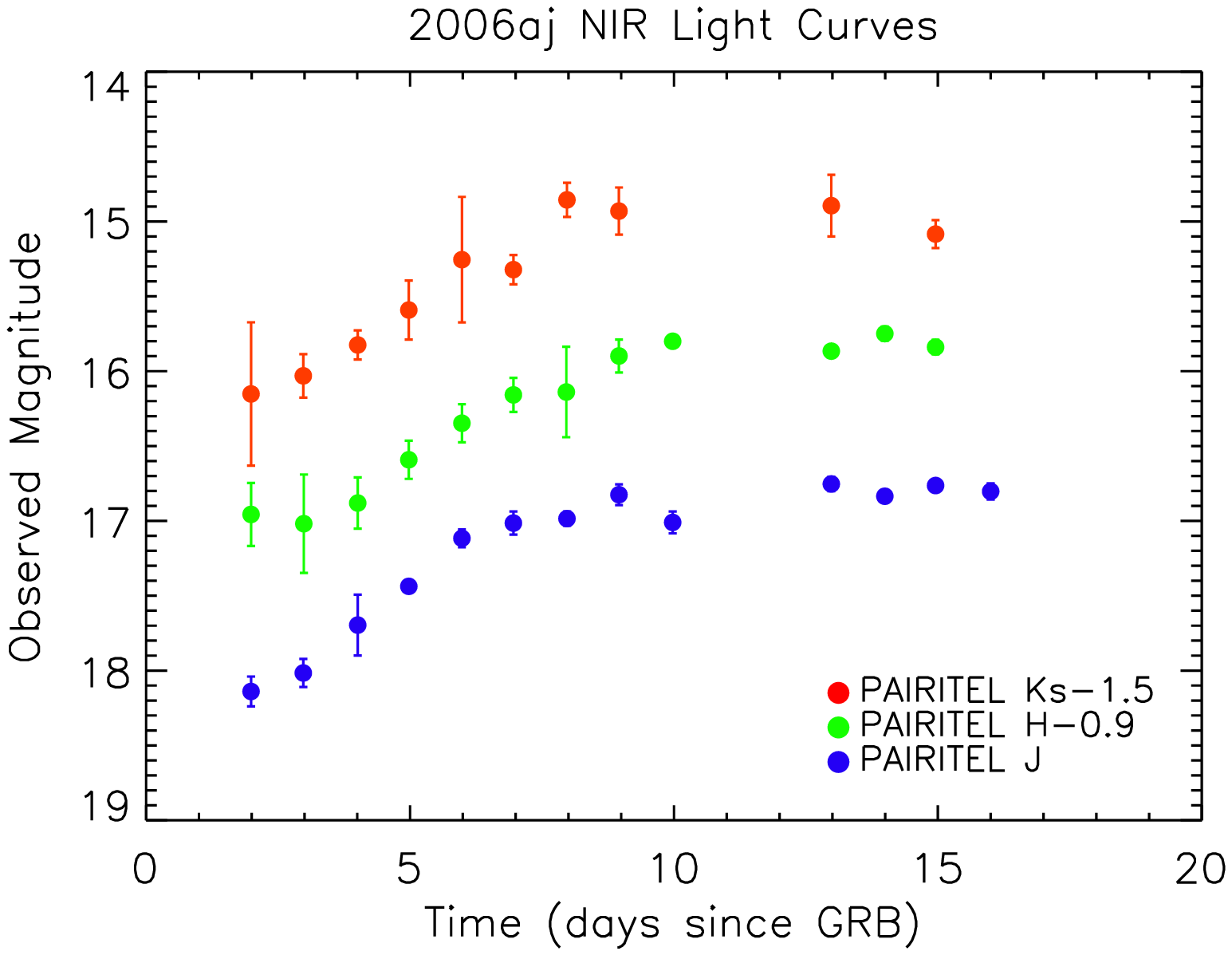}
\end{figure}


\begin{figure}  \label{fig:UBVRIJHKs}
\plotone{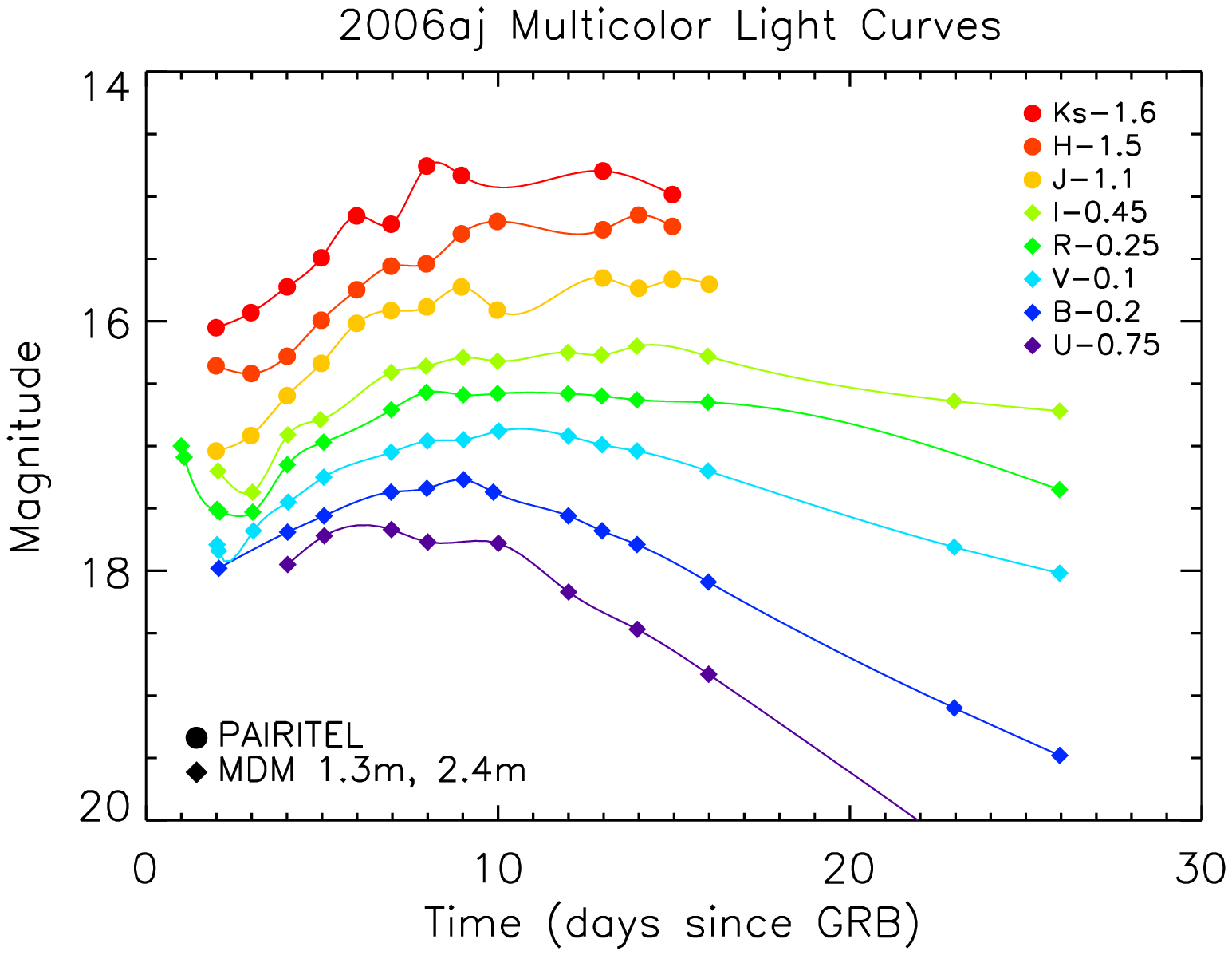}
\end{figure}

\begin{figure}  \label{fig:color-color}
\plotone{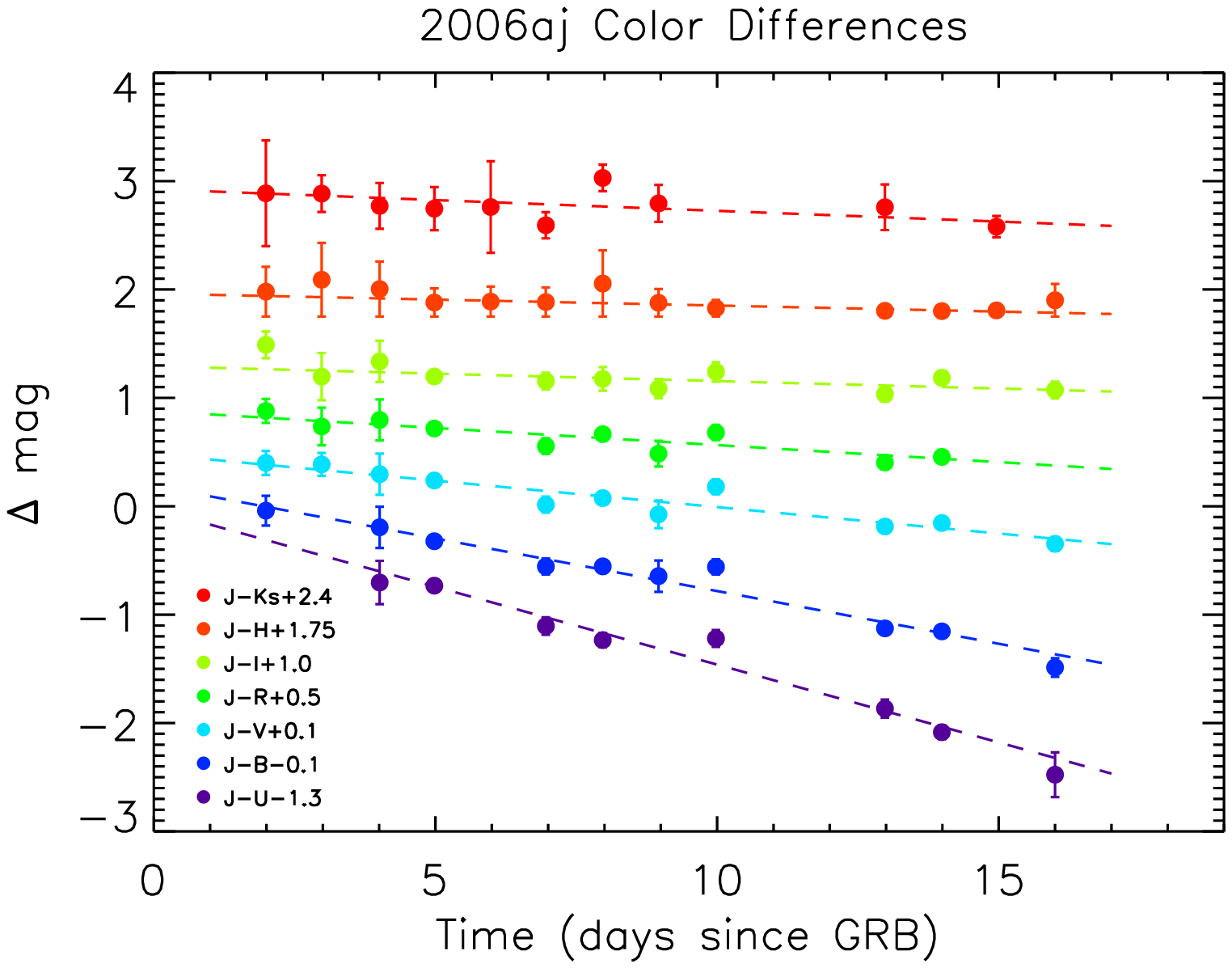}
\end{figure}

\begin{figure}  \label{fig:UBVRIJHKs-SED}
\plotone{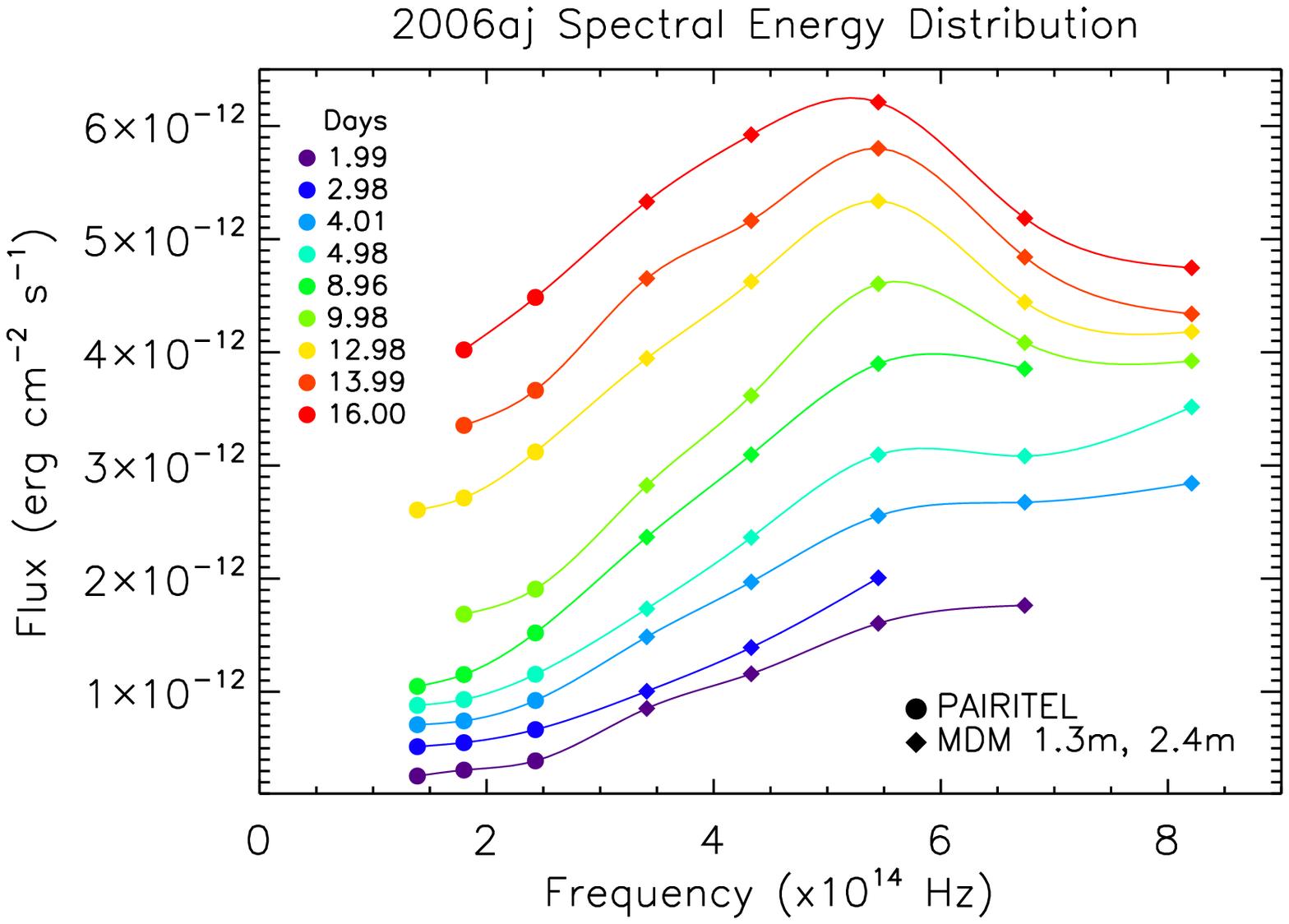}
\end{figure}

\begin{figure}  \label{fig:1998bw}
\plotone{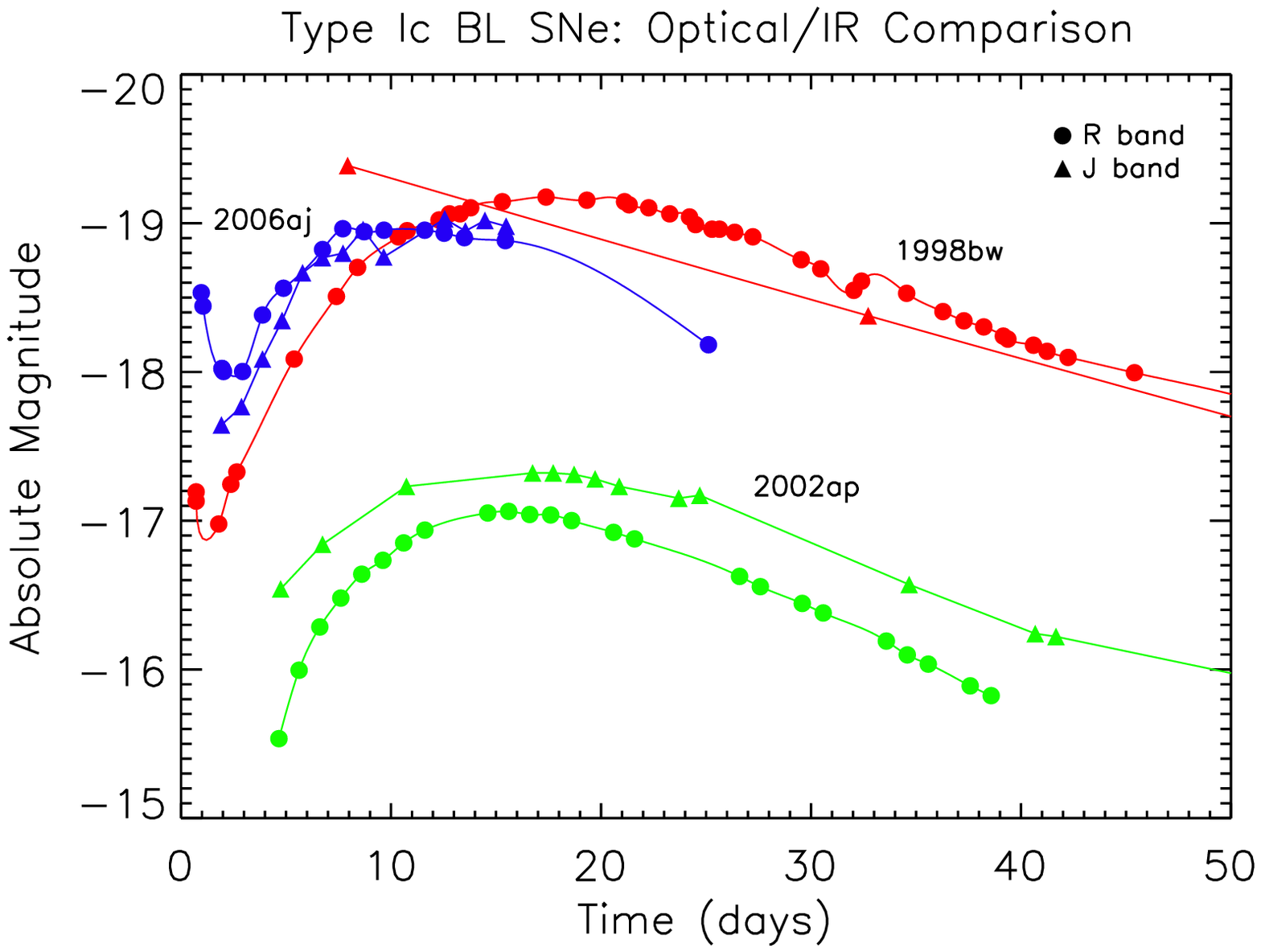}
\end{figure}

\clearpage


 \begin{deluxetable}{rrrrrrrrr}

\label{tbl:data} \tablecolumns{7} \tablewidth{0pc}
\tablecaption{Host Subtracted Photometry of XRF~060218/SN~2006aj }
\tablehead{ \colhead{Days} & \colhead{{\em J} band} & \colhead{{\em H} band}  & \colhead{{\em K$_s$} band} & \colhead{Seeing} & \colhead{Air Mass}  
\\
\colhead{since GRB\tablenotemark{1} } & \colhead{mag \tablenotemark{2} } & \colhead{mag \tablenotemark{2} }  & \colhead{mag \tablenotemark{2} } & \colhead{(arcsec)} & \colhead{(sec z)}  }

\startdata
     
1.99	&	18.31	$\pm$	0.09	&	17.97	$\pm$	0.21	&	17.72	$\pm$	0.48	&	2.93	&	1.10	\\
2.98	&	18.18	$\pm$	0.09	&	18.04	$\pm$	0.33	&	17.26	$\pm$	0.15	&	2.64	&	1.11	\\
4.01	&	17.85	$\pm$	0.19	&	17.89	$\pm$	0.17	&	17.10	$\pm$	0.10	&	2.76	&	1.20	\\
4.98	&	17.58	$\pm$	0.02	&	17.59	$\pm$	0.13	&	16.91	$\pm$	0.20	&	2.77	&	1.22	\\
5.98	&	17.25	$\pm$	0.05	&	17.34	$\pm$	0.13	&	16.63	$\pm$	0.42	&	2.89	&	1.32	\\
6.96	&	17.15	$\pm$	0.07	&	17.15	$\pm$	0.11	&	16.69	$\pm$	0.10	&	2.85	&	1.32	\\
7.97	&	17.12	$\pm$	0.04	&	17.13	$\pm$	0.30	&	16.28	$\pm$	0.11	&	2.92	&	1.43	\\
8.96	&	16.96	$\pm$	0.06	&	16.89	$\pm$	0.11	&	16.35	$\pm$	0.16	&	2.95	&	1.29	\\
9.98	&	17.14	$\pm$	0.07	&	16.79	$\pm$	0.04	&	É		É	&	2.85	&	1.29	\\
12.98	&	16.88	$\pm$	0.04	&	16.85	$\pm$	0.03	&	16.32	$\pm$	0.21	&	1.45	&	1.21	\\
13.99	&	16.97	$\pm$	0.02	&	16.73	$\pm$	0.05	&	É		É	&	2.82	&	1.22	\\
14.96	&	16.89	$\pm$	0.03	&	16.83	$\pm$	0.05	&	16.48	$\pm$	0.09	&	2.64	&	1.39	\\
16.00        &	16.93	$\pm$	0.05	&			&	É	É	&	2.83	&	1.37	\\
     
\enddata

\tablenotetext{1}{Observer frame}
\tablenotetext{2}{Uncorrected for extinction}

\end{deluxetable}

\begin{deluxetable}{rrrrrrrrr}
\label{tbl:data} \tablecolumns{7} \tablewidth{0pc}
\tablecaption{XRF~060218/SN~2006aj Properties }
\tablehead{ \colhead{Property} & \colhead{{\em J} band} & \colhead{{\em H} band}  & \colhead{{\em K$_s$} band} 
\\
\colhead{} & \colhead{mag\tablenotemark{1} } & \colhead{mag\tablenotemark{1}}  & \colhead{mag\tablenotemark{1}} 
}

\startdata

Peak mag & 16.76 $\pm$ 0.03 & $\lesssim$ 16.65 $\pm$ 0.05 & $\lesssim$ 16.39 $\pm$ 0.21 \\
Peak Mag\tablenotemark{2} & $-$19.02 $\pm$ 0.03 &  $\lesssim$ $-$19.13 $\pm$ 0.05 &  $\lesssim$ $-$19.39 $\pm$ 0.24 \\
Host mag	& 18.99 $\pm$ 0.16 & 18.52 $\pm$ 0.22 & 18.69 $\pm$ 0.34 \\
$A_{galactic}$	& 0.112 $\pm$ 0.003 & 0.071 $\pm$ 0.002 & 0.045 $\pm$ 0.001 \\
$A_{host}\tablenotemark{3}$ & 0.035 $\pm$ 0.003 & 0.022 $\pm$ 0.002 & 0.015 $\pm$ 0.001 \\
\enddata

\tablenotetext{1}{Host subtracted and extinction corrected values}
\tablenotetext{2}{$z = 0.0335$, $H_{0} = 72$ km s$^{-1}$, $\Omega_{m} = 0.3$, $\Omega_{\Lambda} = 0.7$ }
\tablenotetext{3}{$R_{V} = 3.1$}

\end{deluxetable}

\begin{deluxetable}{rrrrrrrrr}
\label{tbl:data} \tablecolumns{7} \tablewidth{0pc}
\tablecaption{XRF~060218/SN~2006aj Absolute Magnitudes }
\tablehead{ \colhead{Days} & \colhead{{\em J} band} & \colhead{{\em H} band}  & \colhead{{\em K$_s$} band}  
\\
 \colhead{since GRB\tablenotemark{1}} & \colhead{mag\tablenotemark{2,}\tablenotemark{3}} & \colhead{mag\tablenotemark{2,}\tablenotemark{3}}  & \colhead{mag\tablenotemark{2,}\tablenotemark{3}} }

\startdata
     
1.92	&	$-$17.64	$\pm$	0.09	&	$-$17.92	$\pm$	0.21	&	$-$18.13	$\pm$	0.49	\\
2.88	&	$-$17.76	$\pm$	0.09	&	$-$17.86	$\pm$	0.33	&	$-$18.25	$\pm$	0.15	\\
3.88	&	$-$18.08	$\pm$	0.19	&	$-$18.00	$\pm$	0.17	&	$-$18.45	$\pm$	0.10	\\
4.81	&	$-$18.34	$\pm$	0.03	&	$-$18.29	$\pm$	0.13	&	$-$18.69	$\pm$	0.22	\\
5.79	&	$-$18.66	$\pm$	0.06	&	$-$18.53	$\pm$	0.14	&	$-$19.02	$\pm$	0.48	\\
6.73	&	$-$18.76	$\pm$	0.08	&	$-$18.72	$\pm$	0.13	&	$-$18.96	$\pm$	0.11	\\
7.71	&	$-$18.79	$\pm$	0.05	&	$-$18.74	$\pm$	0.33	&	$-$19.42	$\pm$	0.14	\\
8.67	&	$-$18.95	$\pm$	0.07	&	$-$18.98	$\pm$	0.12	&	$-$19.35	$\pm$	0.19	\\
9.65	&	$-$18.77	$\pm$	0.07	&	$-$19.08	$\pm$	0.04	&				\\
12.56	&	$-$19.02	$\pm$	0.05	&	$-$19.01	$\pm$	0.03	&	$-$19.38	$\pm$	0.24	\\
13.54	&	$-$18.94	$\pm$	0.02	&	$-$19.13	$\pm$	0.05	&				\\
14.47	&	$-$19.01	$\pm$	0.03	&	$-$19.04	$\pm$	0.06	&	$-$19.19	$\pm$	0.11	\\	
     
\enddata

\tablenotetext{1}{Rest frame of GRB}
\tablenotetext{2}{Host subtracted and extinction corrected values}
\tablenotetext{3}{$z = 0.0335$, $H_{0} = 72$ km s$^{-1}$, $\Omega_{m} = 0.3$, $\Omega_{\Lambda} = 0.7$ }

\end{deluxetable}


\end{document}